\documentclass[12pt,preprint]{aastex}
\begin{document}

\title{Probing the Inner Jet of the Quasar PKS~1510$-$089 with Multi-waveband Monitoring during
Strong Gamma-ray Activity}

\author{Alan P. Marscher\altaffilmark{1}, Svetlana G. Jorstad\altaffilmark{1,2},
Valeri M. Larionov\altaffilmark{2,3}, Margo F. Aller\altaffilmark{4}, Hugh D. Aller\altaffilmark{4},
Anne L\"ahteenm\"aki\altaffilmark{5}, Iv\'an Agudo \altaffilmark{6}, Paul S. Smith\altaffilmark{7},
Mark Gurwell\altaffilmark{8}, Vladimir A. Hagen-Thorn\altaffilmark{2,3},
Tatiana S. Konstantinova\altaffilmark{2}, Elena G. Larionova\altaffilmark{2},
Liudmila V. Larionova\altaffilmark{2}, Daria A. Melnichuk\altaffilmark{2},
Dmitry A. Blinov\altaffilmark{2}, Evgenia N. Kopatskaya\altaffilmark{2}, Ivan S.Troitsky\altaffilmark{2},
Merja Tornikoski\altaffilmark{5}, Talvikki Hovatta\altaffilmark{5}, Gary D. Schmidt \altaffilmark{7},
Francesca D. D'Arcangelo\altaffilmark{1,9}, Dipesh Bhattarai\altaffilmark{1}, Brian Taylor\altaffilmark{1,10},
Alice R. Olmstead\altaffilmark{1}, Emily Manne-Nicholas\altaffilmark{1}, Mar Roca-Sogorb\altaffilmark{6}, 
Jos\'e L. G\'omez \altaffilmark{6}, Ian M. McHardy\altaffilmark{11}, Omar Kurtanidze\altaffilmark{12},
Maria G. Nikolashvili\altaffilmark{12}, Givi N. Kimeridze\altaffilmark{12}, and
Lorand A. Sigua\altaffilmark{12}}
\altaffiltext{1}{Institute for Astrophysical Research, Boston University, 725 Commonwealth Avenue, Boston, MA 02215}
\email{marscher@bu.edu}
\altaffiltext{2}{Astronomical Institute, St. Petersburg State University, Universitetskij Pr. 28, Petrodvorets, 
198504 St. Petersburg, Russia}
\altaffiltext{3}{Isaac Newton Institute of Chile, St. Petersburg Branch, St. Petersburg, Russia}
\altaffiltext{4}{Astronomy Department, University of Michigan, 830 Dennison, 500 Church St., Ann Arbor, 
Michigan 48109-1042}
\altaffiltext{5}{Mets\"ahovi Radio Observatory, Helsinki University of Technology TKK, Mets\"ahovintie 114,
FIN-02540 Kylm\"al\"a, Finland}
\altaffiltext{6}{Instituto de Astrof\'{\i}sica de Andaluc\'{\i}a, CSIC, Apartado 3004, 18080,
Granada, Spain}
\altaffiltext{7}{Steward Observatory, University of Arizona, Tucson, AZ 85721-0065}
\altaffiltext{8}{Harvard-Smithsonian Center for Astrophysics, 60 Garden St., Cambridge, MA 02138}
\altaffiltext{9}{Current address: MIT Lincoln Laboratory, 244 WoodSt., Lexington, MA, 02421}
\altaffiltext{10}{Lowell Observatory, Flagstaff, AZ 86001}
\altaffiltext{11}{Department of Physics and Astronomy, University of Southampton, Southampton, SO17 1BJ,
United Kingdom}
\altaffiltext{12}{Abastumani Astrophysical Observatory, Mt. Kanobili, Abastumani, Georgia}

\begin{abstract}

We present results from monitoring the multi-waveband flux, linear polarization, and parsec-scale
structure of the quasar PKS~1510$-$089, concentrating on eight major $\gamma$-ray flares that
occurred during the interval 2009.0-2009.5. The $\gamma$-ray peaks were
essentially simultaneous with maxima at optical wavelengths, although the flux ratio of the two
wavebands varied by an order of magnitude. 
The optical polarization vector rotated by $720^\circ$ during a 5-day period encompassing six
of these flares. This culminated in a very bright, $\sim1$ day, optical and $\gamma$-ray
flare as a bright knot of emission passed through the highest-intensity, stationary feature (the ``core'')
seen in 43 GHz Very Long Baseline Array images. The knot continued
to propagate down the jet at an apparent speed of $22c$ and emit strongly at $\gamma$-ray energies
as a months-long X-ray/radio outburst
intensified. We interpret these events as the result of the knot following a spiral path through a
mainly toroidal magnetic field pattern in the acceleration and collimation zone of the jet, after which it
passes through a standing shock in the 43~GHz core and then continues downstream.
In this picture, the rapid $\gamma$-ray flares
result from scattering of infrared seed photons from a relatively slow sheath of the jet as well as
from optical synchrotron radiation in the faster spine. The 2006-2009.7 radio and X-ray flux variations are
correlated at very high significance; we conclude that the X-rays are mainly from inverse
Compton scattering of infrared seed photons by 20-40 MeV electrons. 
\end{abstract}

\section{Introduction}   
Attempts to understand relativistic jets of blazars have been greatly advanced by the availability
of instruments such as the Very Long Baseline Array (VLBA), the {\it Rossi} X-ray Timing Explorer (RXTE),
and, most recently, the {\it Fermi} Gamma-ray Space Telescope, together with
more traditional telescopes. Long-term multi-waveband monitoring of a number of blazars
with these facilities is now providing valuable insights into the physical processes in the jets
\citep[e.g.,][]{mar08,Chat08,Lar08}. This paper presents a rich set of observations of
the $\gamma$-ray bright quasar PKS~1510$-$089 ($z=0.361$), whose jet exhibits apparent motions of
emission features that are among the fastest (as high as $45c$) of all blazars observed thus far
\citep{jor05}. We analyze motions of features in the parsec-scale radio jet alongside variability of the
optical polarization and flux from radio through $\gamma$-ray frequencies. The relative timing of
correlated variations probes the structure and physics of the innermost jet regions where the flow is
accelerated and collimated, and where the emitting electrons are energized.

\section{Observations and Data Analysis}

Our observations include imaging with the VLBA at 43~GHz
in both total and linearly polarized intensity, with angular resolution
near 0.10~milliarcseconds (mas), or a projected distance of 0.50~pc
\citep[for $H_\circ=71$ km~s$^{-1}$~Mpc$^{-1}$ and the concordance cosmology;][]{spe07}.
We processed the data and created images in a manner identical to that described by
\citet{jor05}.

We derived 0.1-200~GeV $\gamma$-ray fluxes
by analyzing data from the Large Area Telescope (LAT) of the {\it Fermi} Gamma-ray Space Telescope
with the standard software \citep{Atw09}. Photon counts from a circular
region of radius $20^\circ$ centered on PKS~1510$-$089 were fit by single power-law spectral
models of this source, PKS~1502+106, and PKS~1508$-$055, plus model~v02 of the
Galactic and extragalactic backgrounds. The mean slope of the photon spectrum for 2009.0-2009.5
for 7-day integrations, $-2.48\pm0.05$, matched that reported by \citet{abdo09} for 2008.6-2008.83.
Our final fluxes are from 1-day integrations using this
slope, with a detection criterion that the maximum-likelihood test statistic
\citep{Mat96} exceed 9.0.

We obtained 2.4-10~keV X-ray fluxes with the RXTE~PCA. We
processed the data as described by \citet{mar08}, fitting the
photon count spectrum with a single power law plus photoelectric absorption corresponding
to a neutral hydrogen column density of $8\times10^{20}$~cm$^{-2}$. In addition, we measured
flux densities at: 14.5~GHz with the 26~m antenna of the
University of Michigan Radio Astronomy Observatory \citep[see][]{all85};
37~GHz with the 14~m telescope of the
Mets\"ahovi Radio Observatory \citep[see][]{ter98}; and frequencies near 230~GHz
with the Submillimeter Array \citep[see][]{gur07}.

We measured the degree of optical linear polarization $P$ and its position angle $\chi$
with: the 0.4~m telescope of St. Petersburg State University; the 0.7~m telescope at the Crimean
Astrophysical Observatory; the PRISM camera on the Lowell Observatory 1.83~m
Perkins Telescope; the Steward Observatory 2.3 and 1.54~m telescopes
\footnote{Data listing:
http://james.as.arizona.edu/$\sim$psmith/Fermi}; and
the 2.2~m telescope at Calar Alto Observatory, under the MAPCAT program.
The data analysis procedures for the various
telescopes are described in \citet{darc07,Lar08,jor10}. We also obtained
optical (R-band) flux densities from
photometric observations at the six telescopes listed above, the 2.0~m Liverpool
Telescope, and the 0.7~m Meniscus Telescope of Abastumani Astrophysical Observatory. We have added
data from Yale University \footnote{Data listing: http://www.astro.yale.edu/smarts/glast}.

\section{Observational Results and Discussion}

\begin{figure}
\epsscale{.80}
\plotone{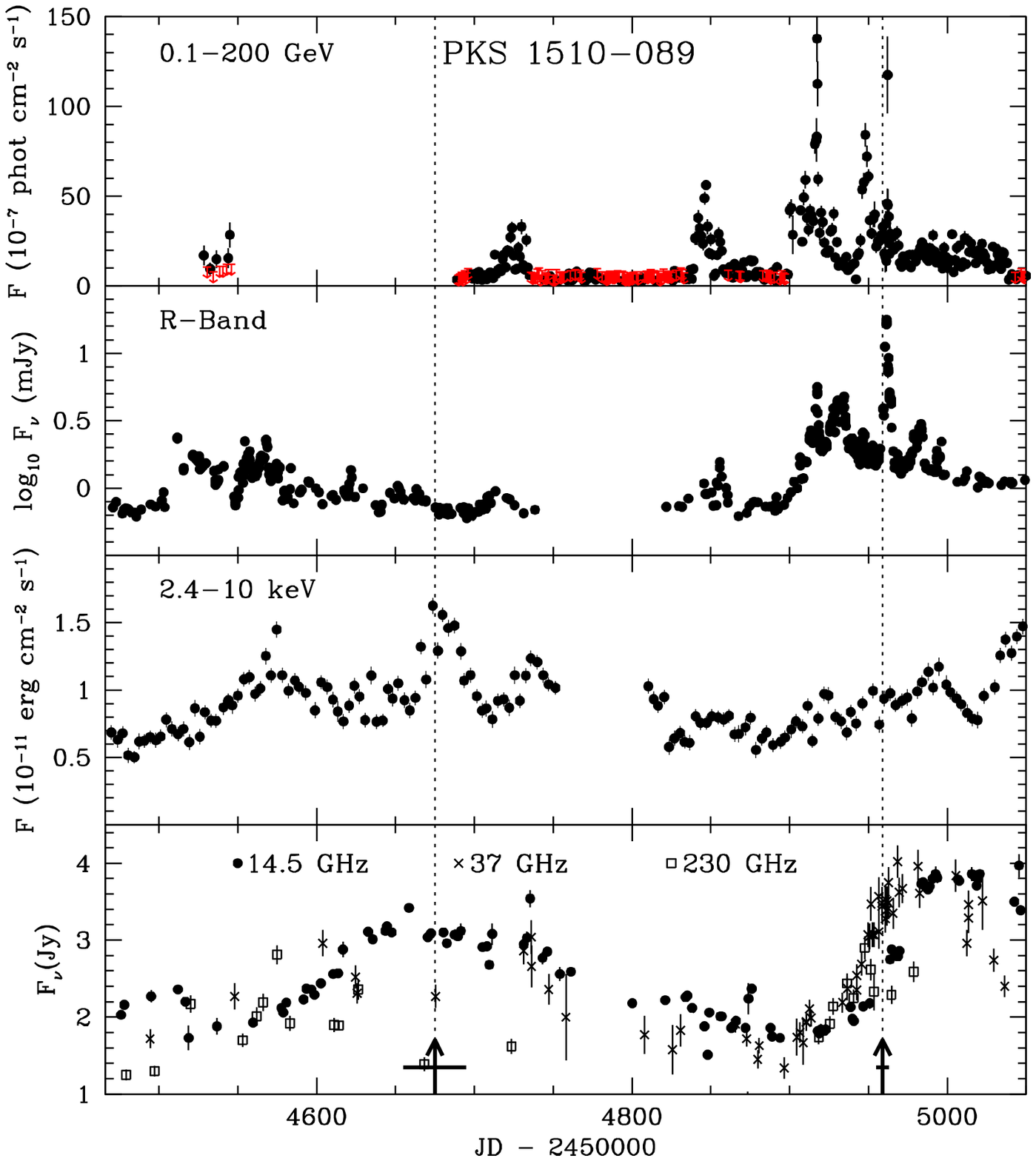}
\caption{Multi-waveband light curves from 2008.0 to 2009.6. Upper limits (red) are 2-$\sigma$.
R-band flux is uncorrected for reddening. Vertical arrows: times when superluminal knots
passed the 43~GHz core; horizontal bars: uncertainties in these times. For reference,
JD~2454700=2008~August~21, 2454900=2009~March~9.}
\label{fig1}
\end{figure}

\begin{figure}
\epsscale{.80}
\plotone{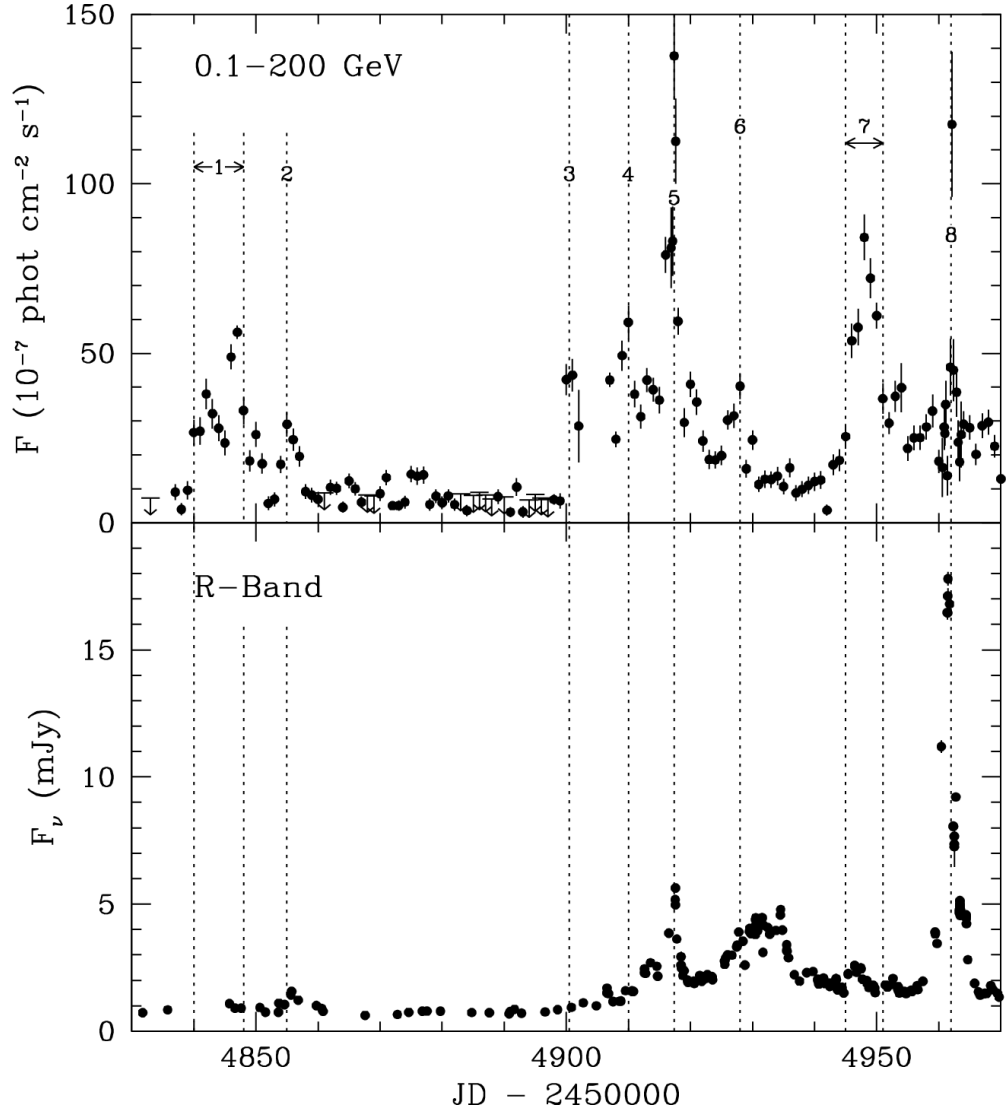}
\caption{$\gamma$-ray and optical light curves from 2009.0 to 2009.38. See caption to
Fig.~\ref{fig1}. Vertical dotted lines denote $\gamma$-ray flares discussed in text.}
\label{fig2}
\end{figure}

\begin{figure}
\epsscale{1.0}
\plotone{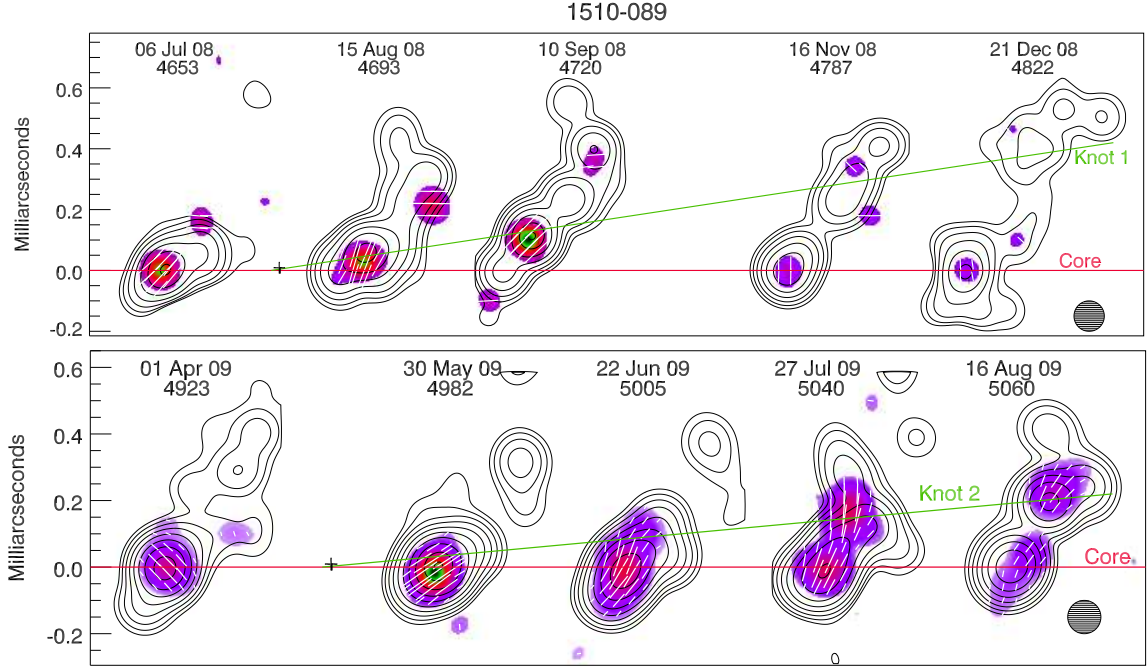}
\caption{Sequence of 43~GHz VLBA images showing ejections of two superluminal
knots with proper motions of (top) $1.1\pm0.1$, (bottom) $0.97\pm0.06$~mas~yr$^{-1}$.
Images are convolved with a circular Gaussian beam
of FWHM=0.1~mas (shaded circle on bottom right). Calendar dates and JD$-2450000$
of images are given. Contours: total intensity, with levels
({\it top}) 1,~2,~4,..., 64,~96\% of peak of 1.41~Jy~beam$^{-1}$, and ({\it bottom})
0.25,~0.5,~1,~2,..., 64,~96\% of peak of 3.14~Jy~beam$^{-1}$. White line segments: direction
of linear polarization; color: polarized intensity relative to peak (green)
of ({\it top}) 71~mJy~beam$^{-1}$ and ({\it bottom}) 120~mJy~beam$^{-1}$.
Polarization not correction for Faraday rotation, whose value is poorly constrained \citep{jor07}.}
\label{fig3}
\end{figure}

A number of flares are apparent in the 2008-09 radio to $\gamma$-ray light curves \citep[Fig.~\ref{fig1},
which includes $\gamma$-ray data from AGILE in 2008 March;][]{Dam09}. Of particular interest is
the first half of 2009 (Fig.~\ref{fig2}), during which eight major $\gamma$-ray flares are apparent.
The VLBA images (Fig.~\ref{fig3}) feature a bright ``core,'' presumed stationary, from which knots
of emission separate at apparent superluminal speeds. We detect two new knots, with apparent speeds
of $24\pm2\,c$ and $21.6\pm0.6\,c$.
The first passed the core on JD~$2454675\pm 20$, as the X-ray flux reached a sharp peak, and
$\sim 2$~weeks after a broad maximum in the 14.5~GHz flux (see Fig.~\ref{fig1}). The second knot passed
the core on JD~$2454959\pm 4$, essentially simultaneous with the extremely sharp, high-amplitude
optical/$\gamma$-ray flare~8 on JD~2454962 (Fig.~\ref{fig2}). We argue below that this
knot was responsible for flares 3-8 and possibly 1-2.

\begin{figure}
\epsscale{.60}
\plotone{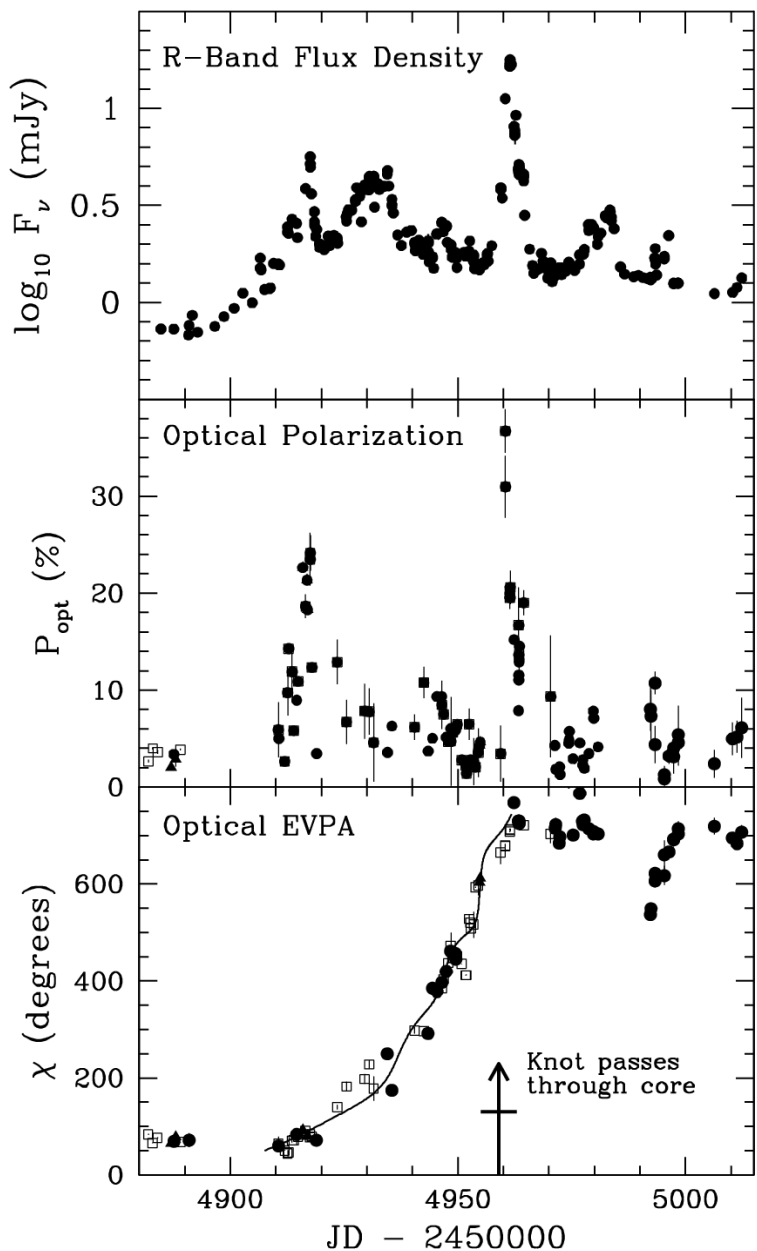}
\caption{R-band flux density and degree~$P$ and electric-vector position angle~$\chi$ of optical polarization
in early 2009. Filled black circles: R-band; filled triangles: V-band; open squares:
$\lambda=$500-700~nm. Multiples of $180^\circ$ are added to $\chi$ as needed to minimize jumps in
consecutive values of $\chi$ or, after JD~2454990, so the values of $\chi$ can be compared with
the end of the first rotation.
The curve fits the $\chi$ data with the model discussed in the text. Highest
amplitude optical flare peaked on JD~2454962=2009~May~10.}
\label{fig4}
\end{figure}

Flare~8 coincides with the end of a 50-day rotation of $\chi$ by $720^\circ$ that started near flare~3
(see Fig.~\ref{fig4}). This striking phenomenon
can be explained either by a stochastic process or by a coherent magnetic field geometry
of the flaring region. In the stochastic interpretation, the magnetic field is turbulent
\citep[see][]{jones88,MGT92,darc09} and the apparent rotation results from a random walk
of the resultant polarization vector direction as cells with random magnetic field orientations
enter and then exit the emission region. According to our
simulations \citep[described by][]{darc07}, one rotation by $> 720^\circ$
similar to that of PKS~1510$-$089 and lasting $50\pm10$ days occurs
once per $\sim 2,000$~days. The probability that such a random apparent rotation would coincide
so closely with a specific 50-day period of elevated $\gamma$-ray flux is extremely small, $\sim0.1\%$.
Furthermore, the model predicts
equally probable clockwise and counterclockwise rotations of $\chi$ in the same object, while the
$\sim180^\circ$ rotation of $\chi$ between JD~2454990 and 2455000 (see Fig.~\ref{fig4})
proceeded at the same counterclockwise rate as during the final stage of the larger rotation. This implies
that the rotation is dictated by geometry, with the accompanying minor flare occurring in the
same location as flare~8.

We observed a similar, but shorter, rotation of the optical polarization in BL Lac in 2005 \citep{mar08}.
We apply our phenomenological model developed to explain that event to PKS~1510$-$089.
A moving emission feature
follows a spiral path as it propagates through the toroidal magnetic field of the acceleration and collimation
zone of the jet \citep{kom07}. The spiral motion is caused either by rotation of the flow \citep{vlah06} or
a helical stream of electron-positron pairs injected from the black hole's ergosphere \citep{rkw04}.
The secular increase in the rate of rotation (see Fig.\ \ref{fig4}) is due to an increasing Doppler
factor as the flow accelerates until it reaches the core, since the Lorentz factor increases as
the cross-sectional radius of the jet \citep{vk04,vlah06,kom07}. [However, the rest-frame angular
velocity of the emission feature cannot decrease to conserve specific angular momentum as the jet expands
\citep{stef95,vlah06}, hence the spiral path followed by the centroid of the feature must maintain a
constant radius of helical motion to avoid this.]
The emission feature must cover most of the cross-section of the jet in order to create substantial
flares and to cancel most of the
polarization (from different orientations of the toroidal magnetic field across the feature) while
leaving a residual of $\lesssim10$\%, with $\chi$ rotating systematically as the feature proceeds down the jet.
Front-to-back light-travel time delays also cancel some polarization by stretching
the feature along the spiral path as viewed by the observer. Synchrotron flares occur
when the energization of electrons increases suddenly over some or all of the emission region; our
model does not attempt to explain these sudden surges in particle acceleration.
The solid curve in the bottom panel of Figure \ref{fig4} shows the fit to the variation of $\chi$ of the
model, with parameter values that are typical but probably not unique. The
bulk Lorentz factor increases linearly with longitudinal distance down the jet from a value of 8 at
the onset of the rotation to 24 at the end. The viewing angle of the jet axis is $1.4^\circ$
in the model, so the Doppler factor increases from 15 to 38 during this time, and the final apparent
speed is $21c$, matching the observed value within the uncertainties. The inflections of the curve are
due to changing
aberration, which we calculate with the equations of \citet{kc85}. The scenario proposed in that paper,
in which the jet twists from helical instabilities, can explain the rotation of $\chi$ but not
the observed low, randomly changing degree of polarization. The secular increase both in the
rate of rotation of $\chi$ and in optical flux during the first
half of the outburst results from the acceleration of the flow and consequent enhanced beaming.

Flare~8, featuring the highest optical flux observed since 1948 \citep{lil75}, occurred as the
rotation of $\chi$ was ending and the new superluminal knot was passing the core. This event
can be explained by compression of the knot by a standing conical shock \citep{mar08}.
According to our model, when the knot passed the core it
was propagating down the jet at 0.3~pc~day$^{-1}$ in our frame, and had traveled 17~pc downstream since
the rotation of $\chi$ began. The transverse radius
of the jet at this point is $\sim 2\times10^{17}$ cm, based on the opening angle of the jet
$\sim 0.2^\circ$ derived by \citet{jor05}. The propagation rate, plus the 1-day timescale of
the flare, sets the longitudinal size of the optical flare region at $\sim 0.3$~pc. We associate
this size with the distance moved by an electron radiating at R-band before it loses too much
energy to continue doing so. To determine the energy loss rate for all of the events,
we first calculate the ratio of
$\gamma$-ray ($> 100$~MeV) to synchrotron luminosity (which equals the ratio of inverse-Compton to
synchrotron loss rate), $\zeta_{\rm gs}$, of the peak of flares 1-8
to be 70, 30, 40, 40, 30, 10, 40, and 9, respectively. We estimate the luminosity
of the synchrotron radiation, whose spectral energy distribution peaks at infrared wavelengths,
as the R-band flux density multiplied by $\nu_R=4.7\times10^{14}$~Hz, by 1.27 to
correct for extinction \citep{Schl98}, and then by 6 to convert roughly to bolometric luminosity
\citep[see][]{kat08}. We then derive that the magnetic field
$B\sim[(6\times 10^6~{\rm s})(\Gamma/20)c/(0.3\xi~{\rm pc})]^{2/3}(\delta/40)^{1/3} \sim 0.4\xi^{-2/3}$~G
during flare~8, where $\xi\equiv(u_{\rm B}+u_{\rm phot})/u_{\rm B}$, $\xi\zeta_{\rm gs}\sim9$
if the seed photons originate from outside the emission feature, $\xi \sim 1$ for mainly first-order SSC, and
$\xi\sim\zeta_{\rm gs}^{1/2}\sim 3$ for combined first- and second-order SSC. We therefore obtain
$B \sim 0.1$--0.4~G, with the highest value for the first-order SSC case.
The (toroidal) magnetic field at the start of the rotation of $\chi$ was then $\sim 1$~G.

The 50-day rotation of $\chi$ implies that a single moving emission feature was responsible for the
entire outburst encompassing flares 3-8. We identify this feature as the superluminal knot seen later
in the VLBA images (Fig.~\ref{fig3}), whose polarization vector lies in the same direction as in
the optical shortly after flare~8. Our observations therefore demonstrate that the high-energy
emission from the jet of PKS~1510$-$089 is quite complex, arising from different regions and probably
by multiple emission mechanisms as a {\it single disturbance} propagates down the jet. Both the $\gamma$-ray
and optical emission is highly variable, but not always in
unison. This is unexpected, since when $B \sim 0.1$--1~G, electrons with the
same energies, $\gamma_e\sim 10^{3.5\pm 0.3}$ in rest-mass units, should be involved in optical
synchrotron radiation and inverse Compton scattering of optical or IR photons to $\gamma$-ray energies.

When the emission feature was close to the base of the jet, electrons in the jet could have scattered the
broad emission-line region or accretion-disk photons to $\gamma$-ray energies as the disturbance first
became optically thin to photon-photon pair production \citep[see, e.g.,][]{gt09,derm09}.
We speculate that this could correspond to flares 1-2.
In this case, the subsequent quiescent period would imply that there is a section of the jet where neither
seed photons nor electrons with $\gamma_e\gtrsim 10^3$ are abundant.

We conclude that some or all of $\gamma$-ray flares~1-4 and 7, with very weak optical counterparts,
were caused by sudden increases in the local seed photon field at optical or
IR wavelengths rather than by increased energization of electrons.
The time scales of variability limit the size of each source of seed photons to $a\lesssim1$~pc---not
much larger than the cross-sectional radius of the jet, $\sim 0.1$~pc---while the luminosity
of each would need to be
$\sim3\times10^{43}(\zeta_{\rm gs}/60)(\Gamma/20)^{-2}(B/0.4~{\rm G})^2(a/0.1$ pc$)^2$erg s$^{-1}$
if the source lies at the periphery of the jet, and higher if more remote.
This luminosity is too high for any commonly occurring cosmic object located parsecs from the central
engine, but could be obtained in
a relatively slow sheath surrounding the ultra-fast spine of the jet responsible for the high
superluminal motion \citep{gtc05}. Moving knots or standing shocks in the sheath could produce the
requisite number of seed photons while being too poorly beamed to contribute substantially to the
observed flux. The relatively slow motion and gradual evolution of such features
implies that they should persist for years, in which case another series of flares in the near
future should exhibit a similar pattern of variability and appearance of a superluminal knot.

Flares~5 and 8 included rapid optical flaring that required sudden energization of electrons
to $\gamma\sim 10^{3.5\pm 0.3}$.
This suggests that the flaring $\gamma$-rays could have been created by the SSC process, an
inference that is supported by the lower value of $\zeta_{\rm gs}$ than for the other flares.
This ratio was, however, much greater than unity, which implies that
second-order scattering contributed significantly to the $\gamma$-ray flux \citep[e.g.,][]{BM96}.
We should therefore expect a
different slope of the $\gamma$-ray photon spectrum for these flares than for flares~1-4 and 7.
There are a sufficient number of photon counts to measure the slopes for flares~5, 7, and 8:
$-2.36\pm 0.09$, $-2.73\pm 0.17$, and $-1.92\pm 0.21$, respectively. The probability that these are
the same is $\sim 0.2\%$. Detailed modeling is needed to determine the slopes expected for the
different radiation mechanisms.

The extremely high optical flux and polarization of flare~8 imply
that the magnetic field was higher and more ordered for this
event than for other flares. This is consistent with our interpretation that the 43~GHz
core---where flare 8 occurred---is a standing shock. The low value of $\zeta_{\rm gs}$
implies that the magnetic field was compressed by a factor $\sim2$ over the previous flare.
We suggest that the emission region responsible for flare~5, when the polarization increased to
$>20\%$, was a subset
of the entire disturbance, so that there was less cancellation of the polarization. This indicates
that the flare was caused by unusually efficient energization of electrons over only part of
the disturbance, although we have insufficient information to determine the cause. In contrast, flares 6
and 7 occurred when the polarization was quite low, which implies that electrons were accelerated less
intensely throughout the entire disturbance.

The flare in 2008 September (JD~2454700-2454750; Fig.~\ref{fig1}) is of a different nature than flares~1-8.
It took place one month after a very bright superluminal knot passed through the core (which coincided with
the peak of a major X-ray flare), consistent with the finding of
\citet{jor01} and \citet{lah03} that high $\gamma$-ray states often follow the time
when a new superluminal knot coincides with the core or a millimeter-wave outburst starts.
The frequency-dependent delays ---
the emission peaked first at optical, then $\gamma$-ray (10-day lag), and then radio and X-ray
(24-day lag) frequencies --- contrast with the simultaneity of the optical and $\gamma$-ray maxima
of flares~2, 5, 7, and 8. The time delays imply that a power law is maintained over only a limited
range of electron energies, with the range changing as a knot separates from the core.

\begin{figure}
\epsscale{.80}
\plotone{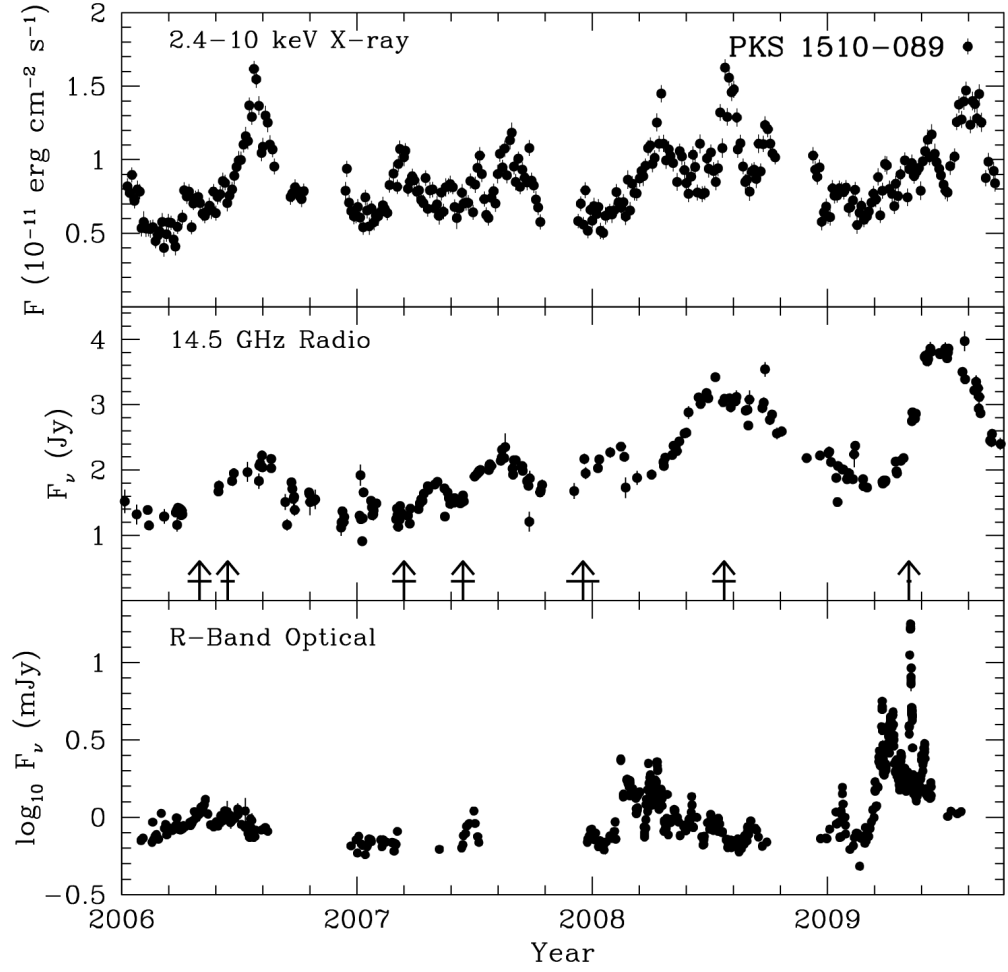}
\caption{Variation of X-ray, radio, and optical flux of PKS~1510$-$089 from 2006.0 to 2009.7. See caption
to Fig.~\ref{fig1}.}
\label{fig5}
\end{figure}

Since 1996, the X-ray emission has correlated better with the 14.5~GHz variations than with those at
higher frequencies \citep{mar06}. During the 2006-2009 period covered by Figure~\ref{fig5},
the correlation is significant at $> 99\%$ confidence. The electrons radiating at 14.5~GHz near the core
have energies $\gamma_e\sim20$-40 in rest mass units, far too small to scatter the
14.5~GHz photons to X-ray energies.
For this reason, we conclude, as have \citet{kat08}, that these electrons produce the X-rays via
inverse Compton scattering of IR seed photons from either the sheath of the jet or a source
outside the jet, such as a dust torus \citep{bla00}. In this case, the X-ray variability is in response
to both the evolution of the electron energy distribution---which affects the radio emission in a
similar way --- and changes in the seed photon density as a knot propagates down
the jet. In the observer's frame, the radiative energy loss time scale of $\gamma_e\sim30$ electrons,
$t_{\rm loss}\sim 3\times10^7 B^{-2}\zeta_{\rm gs}^{-1}\delta^{-1}$~s,
$\sim 5$~days when $B\sim0.4$~G and $\delta\zeta_{\rm gs}\sim400$. This is commensurate with
the fastest X-ray and radio variability that we have observed.

\section{Conclusions}

Our comprehensive dataset has revealed some patterns that allow us to probe the locations and causes
of the high-energy emission in PKS~1510$-$089.
The rotation of the optical polarization vector over 50 days in 2009, as multiple $\gamma$-ray
and optical flares took place, implies that a single emission feature---the superluminal knot later
seen in our VLBA images---was responsible for these
variations. As for BL~Lac \citep{mar08}, we model this as a structure that followed a spiral path
through a helical magnetic field where the jet flow accelerates. The $\gamma$-ray to synchrotron
flux ratio varied greatly among the different flares. This
requires local sources of seed photons from both within the jet and just outside, probably in a
surrounding sheath. As a consequence, flares erupt at a variety of locations as disturbances pass down
the jet. Future comprehensive multi-waveband monitoring plus VLBA imaging
will determine the extent to which our findings apply to the general population of blazars.

\acknowledgments
Funding of this research included NASA grants NNX08AJ64G and NNX08AV65G and NNX08AV65G,
NSF grants AST-0907893 and AST-0607523 (U.~Michigan), Russian RFBR grant~09-02-0092,
Spanish ``Ministerio de Ciencia e Innovaci\'on'' grant AYA2007-67626-C03-03, the Academy of
Finland, the University of Michigan, and Georgian National Science Foundation grant GNSF/ST08/4-404.
The VLBA is an instrument of the National Radio Astronomy Observatory, a facility of the NSF,
operated under cooperative agreement by Associated Universities, Inc. The Calar Alto
Observatory is jointly operated by the Max-Planck-Institut f\"ur Astronomie and the Instituto de 
Astrof\'{\i}sica de Andaluc\'{\i}a-CSIC. The SMA is a joint project between the Smithsonian 
Astrophysical Observatory and the Academia Sinica Institute of Astronomy and Astrophysics, funded by
the Smithsonian Institution and the Academia Sinica.
The Liverpool Telescope, operated on the island of La Palma by Liverpool John Moores University in the
Spanish Observatorio del Roque de los Muchachos of the Instituto de Astrofisica de Canarias, is funded by
the UK Science and Technology Facilities Council.

{\it Facilities:} VLBA, RXTE, Fermi, Liverpool:2m, Perkins, Steward:2.3m,1.54m, Calar Alto:2.2m, UMRAO, SMA

\clearpage

\clearpage

\clearpage

\clearpage

\clearpage

\end{document}